\documentstyle[preprint,aps,eqsecnum]{revtex}
\tighten
\begin{document}
\title {On Orbital Angular Momentum in Deep Inelastic Scattering}

\author{A. Harindranath and Rajen Kundu}
\address{ Saha Institute of Nuclear Physics \\
Sector I, Block AF, Bidhan Nagar,
Calcutta 700064 India}
\date{November 18, 1998}
\maketitle
\begin{abstract}
In this work we address several issues associated with orbital 
angular momentum 
relevant for leading twist polarized deep inelastic scattering.
We present a detailed analysis of the
light-front helicity operator (generator of rotations in the transverse
plane) in QCD.  
We explicitly show that,
the operator constructed from the manifestly gauge
invariant, symmetric energy momentum tensor in QCD, in the gauge $A^+=0$,
after the elimination of constraint variables, is equal to the naive
canonical form of the light-front helicity operator plus surface terms.
Restricting to topologically trivial sector, we eliminate the residual gauge
degrees of freedom and surface terms. Having constructed the gauge fixed
light-front helicity operator,  
we introduce quark and gluon orbital helicity 
distribution functions relevant for polarized deep inelastic scattering as 
Fourier transform of the forward hadron matrix elements of appropriate bilocal 
operators. The utility of these definitions is  illustrated with the calculation 
of anomalous dimensions in perturbation theory. We explicitly verify the
helicity sum rule for dressed quark and gluon targets in light-front perturbation
theory. 
We also consider  
internal orbital helicity of a composite system in an arbitrary reference
frame and contrast the results in the non-relativistic situation versus the 
light-front (relativistic) case.           
\end{abstract}
\vskip .2in
\centerline{PACS: 11.10Ef; 11.40.-q; 12.38.Bx}
\section {Introduction}
The role of orbital angular momentum in deep inelastic scattering was  
first emphasized by Sehgal\cite{lms} and Ratcliff\cite{pgr}.
Recently, orbital angular momentum in QCD has attracted lot of attention
\cite{ji} in
the context of the composition of nucleon helicity (in light-front
quantization) in terms of quark and gluon degrees of freedom. The well-known
polarized structure function $g_1$ measures (ignoring anomaly) the chirality 
of quarks and anti-quarks in the nucleon in the deep inelastic region.
On the light-front, chirality coincides with helicity and thus $g_1$
constitutes a measurement of the intrinsic helicity of fermionic
constituents in the nucleon. Since experimentally this contribution to the
nucleon helicity is shown to be very small, great interest has arisen in the
contributions of intrinsic gluon helicity and quark and gluon
internal orbital helicities in understanding the nucleonic spin structure.

In this work, first, we study the generator of rotations in the transverse
plane  (which we call the light-front helicity operator)
in light-front QCD starting from the manifestly
gauge invariant, symmetric, energy-momentum tensor.  
This operator appears to be interaction dependent.
Further, it contains unphysical
degrees of freedom (constraint variables in light-front theory) 
and it is even unclear whether this
operator will generate the correct transformation laws pertaining to an
angular momentum operator. To proceed, we pick the light-front
gauge $A^+=0$ and use the constraint equations to eliminate the unphysical
degrees of freedom. At this stage, we still have the residual gauge freedom
in this gauge associated with $x^-$ independent gauge transformations. We
restrict our considerations to the {\it topologically trivial sector}
 and require
that the physical field vanish at the boundary ($x^{-,\perp} \rightarrow
\infty$). This {\it eliminates the residual gauge freedom} and also all the
{\it surface terms}. The resulting {\it gauge fixed helicity operator} agrees 
with the naive canonical one (which is free of interactions and naturally
separates into quark and gluon orbital and spin helicities) at the
operator level.        

Having constructed the gauge fixed orbital helicity operator
in the topologically
trivial sector of QCD, we proceed to define
non-perturbative parton  distributions 
as the Fourier transform along the light-front longitudinal direction of the
forward hadron matrix element of appropriate bilocal operators for
light-front internal orbital helicity. 
Apart from
providing non-perturbative information on the distribution of nucleon helicity
among its partonic constituents, they serve another useful purpose for the 
perturbative region. As we have recently shown \cite{hari}, by replacing the
hadron target by a dressed parton target in the definition of the
distribution function, one can easily calculate the
splitting functions and corresponding anomalous dimensions of leading twist
operators. One can also investigate other issues of perturbative concern in
the case of higher twist operators. The method uses intuitive light-front 
Fock space expansion for the operators in the bilocal expressions and also for 
the state. We proceed to explicitly evaluate the splitting functions and
corresponding anomalous dimensions relevant for
orbital helicity at one loop level in a physically transparent manner. 

There has been recent studies of the quark orbital motion using
non-relativistic quark models \cite{chengli}. It is of importance to study 
in what respect
the physics of orbital helicity in the relativistic case differs
from its non-relativistic counterpart. In the context of {\it 
gauge-fixed light-front helicity operator}
in QCD which is free from interactions
we address this issue in an appendix.           
\section{Light-front Helicity operator $J^3$ from the manifestly gauge 
invariant energy momentum tensor}
It is well-known that even though the {\it energy-momentum density} (which gives rise to
Hamiltonian and three-momentum) and the {\it generalized angular momentum 
density} (which gives
rise to angular momentum and boosts) can be expressed in a manifestly
covariant, gauge invariant form, the explicit form of Poincare generators 
in quantum 
field theory depends on the frame of reference and may also depend up on the 
gauge choice. This of course does not imply that the 
theory has lost Lorenz and gauge symmetry. The symmetries are no longer 
manifest, but the physical observables in the theory still obey the consequences 
of the symmetries. 

Poincare generators can be further classified as kinematical (which do not
contain interactions and do not change the quantization surface) and
dynamical (which contain interactions and change the quantization surface).
Which operator is dynamical and which is kinematical of course depends on
the choice of quantization surface. It is well-known that in light-front
field theory, on which our formalism of deep inelastic scattering is based,
the generators of boosts and the rotation in the transverse plane
(light-front helicity) are kinematical like three momenta
whereas the generators of rotations
about the two transverse axes are dynamical like the Hamiltonian. 
The operator in light-front
field theory relevant to the $``$proton spin crisis " is the light-front
helicity operator which belongs to the kinematical subgroup. In light-front
literature, it is customary to construct this operator from the canonical
symmetric energy momentum tensor and one explicitly finds that this operator
is indeed free of interaction and has the same form as in free field
theory\cite{ks}.

In non-Abelian gauge theories like QCD, one should be extra cautious since
such theories are known to exhibit non-trivial topological effects. In this
work,  we restrict our attention to the topologically trivial sector of QCD. 
In this sector, interactions do not affect kinematical 
generators\cite{wein}. 
In view of the prevailing confusion in the literature (see Ref. \cite{mhs}
for a list of recent papers on the subject),
we provide an explicit demonstration of this
fact in this section in the case of the light-front helicity operator.
      
We start from the manifestly gauge invariant, symmetric energy momentum
tensor in QCD.
\begin{eqnarray}
\Theta^{\mu \nu} && = { i \over 2} {\overline \psi} [ \gamma^\mu D^\nu +
\gamma^\nu D^\mu ]\psi - F^{\mu \lambda a} F^{\nu a} _{~ \lambda} \nonumber \\
&&  ~~ - g^{\mu \nu} \Big \{ -{1 \over 4} (F_{\lambda \sigma a})^2 +
{\overline \psi}(i \gamma^\lambda D_\lambda -m) \psi \Big \}
\end{eqnarray}
where $ i D^\mu = i \partial^\mu + g A^\mu $, 
$ F^{\mu \lambda a} = \partial^\mu  A^{\lambda a} - \partial^\lambda
A^{\mu a} + g f^{abc} A^{\mu b} A^{\lambda c}$ , $ 
F^{\nu a}_{~ \lambda } = \partial^{\nu a}  A_{\lambda} - \partial_{\lambda}
A^{\nu a} + g f^{abc} A^{\nu b} A^{\lambda c}$.

We define the light-front helicity operator
\begin{eqnarray}
{\cal J}^3 = { 1 \over 2} \int dx^- d^2 x^\perp  [ x^1  \Theta^{+2} - x^2
\Theta^{+1}].
\end{eqnarray}
${\cal J}^3$ is a manifestly gauge invariant operator 
by construction. However, 
it depends explicitly on the interaction and does not appear to be a
kinematical operator at all. Furthermore, 
it is not apparent that ${\cal J}^3$
generates the correct transformations as an angular momentum operator. 
Thus at this stage, we are not justified to call it a helicity operator.

Explicitly, we have,
\begin{eqnarray}
{\cal J}^3 && = { 1 \over 2} \int dx^- d^2 x^\perp \Big \{
x^1  [ { i \over 2} {\overline \psi} (\gamma^+ D^2 + \gamma^2 D^+) \psi 
        - F^{+ \lambda a} F^{2a}_{~\lambda}]  \nonumber \\
&& ~~ - x^2  [ { i \over 2} {\overline \psi} (\gamma^+ D^1 + \gamma^1 D^+) \psi 
        - F^{+ \lambda a} F^{1 a}_{~\lambda}] \Big \}    
\end{eqnarray}
The fermion field can be decomposed as $ \psi^{\pm} = \Lambda^{\pm} \psi$, with $ \Lambda^{\pm} = { 1 \over
4} \gamma^{\mp} \gamma^{\pm}$. 
We shall work in the gauge $A^+=0$. 
In this gauge, we still have residual gauge freedom associated with
$x^-$-independent gauge transformations. Note that only $\psi^+$ and $A^i$
are dynamical variables whereas $\psi^-$ and $A^-$ are constrained.

We have,
\begin{eqnarray}
{ i \over 2} {\overline \psi} ( \gamma^+ D^2  + \gamma^2 D^+) \psi =
{\psi^+}^\dagger i \partial^2 \psi^+  
+ g {\psi^+}^\dagger T^a \psi^+ A^2_a + { i \over 2} {\overline \psi}
\gamma^2 i \partial^+ \psi.
\end{eqnarray}
Using the constraint equation
\begin{eqnarray}
i \partial^+ \psi^- = \Big [ \alpha^\perp \cdot  ( i \partial^\perp + g
A^\perp) + \gamma^0 m \Big ] \psi^+,
\end{eqnarray}
to eliminate the constraint variable $\psi^-$ we arrive at, 
after some algebra,
\begin{eqnarray}
{ i \over 2} {\overline \psi} \gamma^2 \partial^+ \psi &&= i {\psi^+}^\dagger
\partial^2 \psi^+ 
+ {1 \over 2} \partial^1({\psi^+}^\dagger \Sigma^3 \psi^+) + g {\psi^+}^\dagger T^a
\psi^+ A^{2a} \nonumber \\
&& ~~~~ + { i \over 2} \partial^+ \Big ( {\psi^-}^\dagger \alpha_2 \psi^+
\Big ) - { i \over 2} \partial^2 \Big ( {\psi^+}^\dagger \psi^+ \Big ).
\end{eqnarray}
Now we restrict ourselves to the topologically trivial sector by
requiring that the dynamical fields ($\psi^+$ and $A^i$) vanish at $x^{-,i}
\rightarrow \infty$. The residual gauge freedom and the surface terms 
are no longer present 
and so we drop total derivatives of
$\partial^+$ and $ \partial^2$. Note that the term involving $ \partial^1$
is not a surface term since $\Theta^{+2}$ is multiplied by $x^1$.  

Collecting the results together, we have,
\begin{eqnarray}
{ i \over 2} {\overline \psi} (\gamma^+ D^2 + \gamma^2 D^+) \psi = 2 i
{\psi^+}^\dagger  \partial^2 \psi^+  
+ {1 \over 2} \partial^1 ({\psi^+}^\dagger \Sigma^3 \psi^+) + 2 g {\psi^+}^\dagger T^a
\psi^+ A^{2a}
\end{eqnarray}
where $ \Sigma^3 = i \gamma^1 \gamma^2$.

In the gauge $A^+=0$, 
\begin{eqnarray}
- F^{+ \lambda a} F^{2 a}_{~ \lambda } && = - { 1 \over 2 } (\partial^+)^2 A^{-a}
A^{2a} + \partial^+ A^{ja} (\partial^2 A^{ja} - \partial^j A^{2a}) 
+ g f^{abc} (\partial^+ A^{ja}) A^{2b} A^{jc} \nonumber \\
&& ~~~~ + { 1 \over 2} \partial^+ \Big (\partial^+ A^{-a}  A^{2a} \Big ).
\end{eqnarray}

We have the constraint equation for the elimination of the variable $A^-$,
\begin{eqnarray}
 { 1 \over 2} (\partial^+)^2 A^{-a} = \partial^+ \partial^i A^{ia} + g
f^{abc} A^{ib} \partial^+ A^{ic} + 2 g {\psi^+}^\dagger T^a \psi^+.
\end{eqnarray}
Thus
\begin{eqnarray}
-F^{+ \lambda a} F^{2a}_{~ \lambda } &&= \partial^i A^{ia} \partial^+ A^{2a} +
\partial^+ A^{ja}(\partial^2 A^{ja} - \partial^j A^{2a}) - 2 g
{\psi^+}^\dagger T^a \psi^+ A^{2a} \nonumber \\
&&~~~~~ + { 1 \over 2} \partial^+ \Big ( \partial^+ A^{-a} A^{2a} \Big )
- \partial^+ \Big ( \partial^i A^{ia} A^{2a} \Big ) \nonumber \\
&& = \partial^+ A^{1a} \partial^2 A^{1a} + \partial^+ A^{2a} \partial^2
A^{2a} + \partial^1 (A^{1a} \partial^+ A^{2a}) \nonumber \\
&& ~~~~+ { 1 \over 2} \partial^+ \Big ( \partial^+ A^{-a} A^{2a} \Big )
- \partial^+ \Big ( \partial^i A^{ia} A^{2a} \Big )
- \partial^+ \Big ( A^{1a} \partial^1 A^{2a} \Big ).
\end{eqnarray}

Collecting the results together,
\begin{eqnarray}
\Theta^{+2} && = 2 i {\psi^+}^\dagger \partial^2 \psi^+ +
{ 1 \over 2} \partial^1({\psi^+}^\dagger \Sigma^3 \psi^+) \nonumber \\
&& ~~ + \partial^+ A^{1a} \partial^2 A^{1a} + \partial^+ A^{2a} \partial^2
A^{2a} + \partial^1 (A^{1a} \partial^+ A^{2a}).
\end{eqnarray}
We have dropped the surface terms at $ x^- = \pm \infty$.
By a similar calculation,
\begin{eqnarray}
\Theta^{+1} && = 2 i {\psi^+}^\dagger \partial^1 \psi^+ -
{ 1 \over 2} \partial^2({\psi^+}^\dagger \Sigma^3 \psi^+) \nonumber \\
&& ~~ + \partial^+ A^{1a} \partial^1 A^{1a} + \partial^+ A^{2a} \partial^1
A^{2a} + \partial^2 (A^{2a} \partial^+ A^{1a})
\end{eqnarray}
>From the above two equations it is clear that $\Theta^{+1}$ and
$\Theta^{+2}$ agree with the free field theory form at the operator level.
This shows that
in light-front quantization, with $A^+=0$ gauge, ${\cal J}^3 = J^3$ (the
naive canonical form independent of interactions) at the
operator level, provided the fields vanish at the boundary. Explicitly,
\begin{eqnarray}
J^3 = J^3_{f(o)}+ J^3_{f(i)}+ J^3_{g(o)}+J^3_{g(o)}
\end{eqnarray}
with 
\begin{eqnarray}
J^3_{f(o)} &&= \int dx^- d^2 x^\perp {\psi^+}^\dagger i ( x^1 \partial^2 -x^2
\partial^1) \psi^+ , \nonumber \\
J^3_{f(i)}&& = { 1 \over 2} \int dx^- d^2 x^\perp {\psi^+}^\dagger
\Sigma^3 \psi^+, \nonumber \\
J^3_{g(o)}&& = {1 \over 2} \int dx^- d^2 x^\perp \Big \{ 
x^1 [\partial^+A^1 \partial^2 A^1 + \partial^+A^2 \partial^2 A^2]
-x^2 [\partial^+A^1 \partial^1 A^1 + \partial^+A^2 \partial^1 A^2]
\Big \}, \nonumber \\
J^3_{g(i)} && = { 1 \over 2} \int dx^- d^2 x^\perp [ A^1 \partial^+ A^2 - 
A^2 \partial^+ A^1 ].
\end{eqnarray}
The color indices are implicit in these equations.

Using canonical commutation relations, we explicitly find that,
\begin{eqnarray}
i \left [ J^{3}_{f(o)}, \psi^{+} (x) \right ] &&= (x^1 \partial^2 - x^2 \partial^1)
\psi^+(x), \label{commu1} \\
i \left  [ J^{3}_{f(i)},\psi^{+}(x) \right ] && = { 1 \over 2} \gamma^1
\gamma^2 \psi^{+}(x),  \label{commu2} \\
i \left [ J^{3}_{g(o)},A^{i}(x) \right ]  && =  (x^1 \partial^2 - x^2 \partial^1) 
A^{i}(x),  \label{commu3} \\
i \left [J^{3}_{g(i)}, A^{i} (x)\right ]  && = - \epsilon_{ij} A^{j} (x) .
\label{commu4}
\end{eqnarray}
Thus these operators do qualify as angular momentum operators (generators of
rotations in the transverse plane) in the theory\cite{ks}.

To summarize,
the helicity operator constructed from manifestly gauge
invariant, symmetric, energy momentum tensor in QCD, in the gauge $A^+=0$,
and after the elimination of constraint variables, is equal to the naive
canonical form of the light-front helicity operator plus surface terms. In
the topologically trivial sector, we can legitimately require the dynamical
fields to vanish at the boundary. This eliminates the residual gauge degrees
of freedom and removes the surface terms. Thus we have a gauge fixed
Poincare generator which we consider in the following sections.

\section{ Orbital helicity distribution functions}
We define the orbital helicity distribution for the fermion
\begin{eqnarray}
\Delta q_L(x, Q^2) = { 1 \over 4 \pi P^+} \int d \eta e^{ -i  \eta x} 
\langle PS \mid \Big [ {\overline \psi}(\xi^-) \gamma^+ i (x^1 \partial^2
- x^2 \partial^1) \psi(0) + h.c.\Big ] \mid PS \rangle \label {fo}
\end{eqnarray}
with $ \eta = {1 \over 2} P^+ \xi^-$. Here $ \mid P S \rangle $ denotes the
hadron state with momentum $P$ and helicity $S$.
 
We define the light-front orbital helicity distribution for the gluon as
\begin{eqnarray}
	\Delta g_L(x,Q^2) &=& {-1  \over 4 \pi P^+} \int d \eta 
		e^{ - i \eta x} \langle PS \mid \Big [ x^1 F^{+ \alpha}(\xi^-) 
		\partial^2 A_\alpha(0)  - x^2 F^{+ \alpha} (\xi^-) \partial^1 
		A_\alpha(0) \Big ] \mid PS \rangle. \label{go}
\end{eqnarray}
These distributions are defined in analogy with the more 
familiar intrinsic helicity distributions for quarks and gluons 
given as follows. 

For the fermion, the intrinsic light-front helicity distribution function 
is given by
\begin{eqnarray}
\Delta q(x,Q^2) = { 1 \over 8 \pi S^+} \int d \eta e^{ - i \eta x} 
\langle PS \mid \Big [ {\overline \psi} (\xi^- ) \gamma^+ \Sigma^3 \psi(0) +
h.c \Big ] \mid PS \rangle \label{fi}
\end{eqnarray}
where $ \Sigma^3 = i \gamma^1 \gamma^2$. This is the same as the chirality
distribution function $g_1$.

For the gluon, the intrinsic light-front helicity distribution is defined
\cite{jgd} as
\begin{eqnarray}
\Delta g(x,Q^2) = -{ i \over 4 \pi (P^+)^2 x} \int d \eta e^{ - i \eta x}
\langle PS \mid F^{+ \alpha} (\xi^-) {\tilde F}^+_{~~\alpha}(0) \mid PS \rangle.
\label {gi}
\end{eqnarray}
The dual tensor 
\begin{eqnarray}
{\tilde F^{\mu \nu}} = { 1 \over 2} \epsilon^{\mu \nu \rho \sigma} F_{\rho
\sigma} ~~~~
{\rm with} ~~~~ \epsilon^{+1-2} = 2.
\end{eqnarray}

Note that the above distribution functions are defined in the
light-front gauge $A^+=0$. In the two-component representation\cite{zhang93}
we have the dynamical fermion field,
\begin{eqnarray}
	\psi_+(x) = \sum_\lambda \chi_\lambda \int {dk^+d^2k_\bot
		\over 2(2\pi)^3 \sqrt{k^+}}\Big(b_\lambda(k)e^{-ikx} + 
		d_{-\lambda}^\dagger(k)e^{ikx} \Big)\label{psi} ,
\end{eqnarray}
and the dynamical gauge field
\begin{eqnarray}
	A_{}^i(x) = \sum_\lambda \int {dk^+d^2k_\bot\over 
		2(2\pi)^3k^+}\Big(\varepsilon^i(\lambda) 
		a_\lambda(k)e^{-ikx} + h.c \Big)\label{ap},
\end{eqnarray}
with
\begin{eqnarray} 
	\Big\{b_\lambda (k), b_{\lambda'}^\dagger(k') \Big\} &=& 
		\Big\{d_\lambda(k), d_{\lambda'}^\dagger{k'} \Big\} 
		= 2(2\pi)^3 k^+\delta (k^+-{k'}^+) \delta^2(k_\bot - k_\bot') 
\delta_{\lambda \lambda'}, \\
	\Big[a_\lambda(k) , a_{\lambda'}^\dagger (k') \Big] &=& 
		2(2\pi)^3 k^+\delta (k^+-{k'}^+) \delta^2(k_\bot - k_\bot')
\delta_{\lambda \lambda'}, 
\end{eqnarray}
and $\chi_\lambda$ is the eigenstate of $\sigma_z$ in the two-component
spinor of $\psi_+$ by the use of the following light-front $\gamma$ 
matrix representation \cite{Zhang95},
\begin{equation}
	\gamma^0 = \left[\begin{array}{cc} 0 & - i \\ i & 0 \end{array}
		\right] ~~, ~~ 
	\gamma^3=  \left[\begin{array}{cc} 0 & i \\ i & 0 \end{array}
		\right] ~~, ~~
 	\gamma^i =  \left[\begin{array}{cc} -i\tilde{\sigma}^i & 0 \\ 
		0 & i\tilde{\sigma}^i \end{array} \right] 
\end{equation}
with $\tilde{\sigma}^1 =\sigma^2, \tilde{\sigma}^2=-\sigma^1$) and 
$\varepsilon^i(\lambda)$ the polarization vector of transverse 
gauge field.

Note that integration of the above distribution functions over $x$ is
directly related to the expectation values of the corresponding helicity
operators as follows.
\begin{eqnarray}
\int_0^1 dx  \Delta q(x,Q^2)=&&~ { 1 \over {\cal N}} \langle PS \mid  J^3_{q(i)} 
 \mid PS \rangle\nonumber\\
\int_0^1 dx  \Delta q_L(x,Q^2)=&&~ { 1 \over {\cal N}} \langle PS \mid
J^3_{q(o)} 
 \mid PS \rangle\nonumber\\
\int_0^1 dx  \Delta g(x,Q^2)=&&~ { 1 \over {\cal N}} \langle PS \mid  J^3_{g(i)} 
 \mid PS \rangle\nonumber\\
\int_0^1 dx  \Delta g_L(x,Q^2)=&&~ { 1 \over {\cal N}} \langle PS \mid
J^3_{g(o)} 
 \mid PS \rangle
\end{eqnarray}
where ${\cal N}=2(2\pi)^3P^+\delta^3(0)$.
\section{Perturbative calculation of anomalous dimensions} 
In this section, we evaluate the internal helicity distribution functions for
a dressed quark in perturbative QCD by replacing the hadron target by a
dressed quark target. We have provided the necessary details of the
calculation which may serve as the stepping stone for more realistic
calculation with meson target. From this simple calculation, we have 
illustrated how easily one can 
extract the relevant splitting
functions and evaluate the corresponding anomalous dimensions. 
Note that, since we are not interested in exhaustive calculation of various
anomalous dimensions and the purpose of this section being illustrative, 
we can safely drop the derivative of delta function in the following
calculations and work explicitly with forward matrix element.

The dressed quark state with fixed helicity can 
be expressed as
\begin{eqnarray}
	|k^+,k_\bot,\lambda \rangle &=& \Phi^{\lambda}(k) b^\dagger_\lambda
		(k)|0\rangle + \sum_{\lambda_1\lambda_2} \int {dk_1^+d^2
		k_{\bot 1}\over \sqrt{2(2\pi)^3 k_1^+}} 
                {dk_2^+d^2k_{\bot 2}\over \sqrt{2 (2\pi)^3k_2^+}} 
     \sqrt{2(2\pi)^3 k^+} \delta^3(k-k_1-k_2) \nonumber \\
	&&~~~~~~~~~~~~~~~~~~~~~~~~~~~~~~~\times \Phi^\lambda
		_{\lambda_1\lambda_2}(k;k_1,k_2)b^\dagger_{\lambda_1}
		(k_1) a^\dagger_{\lambda_2} (k_2) | 0 \rangle + 
		\cdots, \label{dsqs}
\end{eqnarray}
where the normalization of the state is determined by
\begin{equation}
	\langle {k'}^+,k'_\bot,\lambda' |k^+,k_\bot,\lambda \rangle
	= 2(2\pi)^3 k^+ \delta_{\lambda,\lambda'}\delta(k^+-{k'}^+)
	\delta^2(k_\bot-k'_\bot), 
\end{equation}
We introduce the boost invariant amplitudes $\psi_1^\lambda$ 
and $ \psi^\lambda_{\sigma_1,
\lambda_2}(x,\kappa^\perp)$ respectively by  $\Phi^\lambda(k)=\psi_1^\lambda$ and 
 $\Phi^\lambda_{\lambda_1\lambda_2}(k;k_1,k_2)  
= { 1 \over \sqrt{P^+}}  \psi^\lambda_{\sigma_1
\lambda_2}(x,\kappa^\perp)$. From the light-front QCD Hamiltonian, to lowest
order in perturbation theory, we have,
\begin{eqnarray}
	\psi^\lambda_{\sigma_1\lambda_2}(x,\kappa_\bot) &=& - 
{g \over \sqrt{2 (2 \pi)^3}}
		T^a {x(1-x)\over \kappa_\bot^2 + m_q^2(1-x)^2}
		\chi^\dagger_{\sigma_1} \Bigg\{2{\kappa_\bot^i \over
		1-x} \nonumber \\
	&& ~~~~~~~~~~~~~ + {1\over x}(\tilde{\sigma_\bot}\cdot \kappa_\bot)
		\tilde{\sigma}^i -im_q\tilde{\sigma}^i{1-x\over x}\Bigg\}
		\chi_\lambda \varepsilon^{i*}(\lambda_2)~\psi_1^\lambda .
 \label{psip}
\end{eqnarray}
Here $x$ is the longitudinal momentum fraction carried by the quark.
We shall ignore the $m_q$ dependence in the above wave function which can
lead to higher twist effects in orbital helicity.
In the following we take the helicity of the dressed
quark to be + $ {1 \over 2}$. Due to transverse boost invariance,
without loss of generality, we take the transverse momentum of the initial
quark to be zero.

First we evaluate the gluon intrinsic helicity
distribution function given in eq. (\ref{gi}) in the dressed quark state.
 
The non vanishing contribution comes from the quark-gluon state. We get,
\begin{eqnarray}
\Delta g (1-x,Q^2) && = \sum_{\sigma_1, \lambda_2} \lambda_2  ~
\int d^2 \kappa^\perp~
{\psi^{\uparrow}_{\sigma_1 \lambda_2}}^*(x, \kappa^\perp) 
{\psi^{\uparrow}_{\sigma_1 \lambda_2}}(x, \kappa^\perp)
\nonumber \\
&& = {\alpha_s \over 2 \pi} C_f ln{Q^2 \over \mu^2} ~x^2 (1-x)^2
{ 1 \over 1-x} \Big [ { 1 \over x^2 (1-x)^2} - { 1 \over (1-x)^2} \Big ].
\end{eqnarray}
The first (second) term inside the square bracket arises from the 
state with gluon helicity +1 (-1).
Thus we have the gluon intrinsic helicity contribution in the dressed quark
state
\begin{eqnarray}
\Delta g(1-x,Q^2) = {\alpha_s \over 2 \pi} C_f ln{Q^2 \over \mu^2}  ~(1+x).
\label{gih}
\end{eqnarray}
Note that the gluon distribution function has the argument $(1-x)$ since we
have assigned $x$ to the quark in the dressed quark state.

Next we evaluate the quark orbital helicity
distribution function given in  eq.(\ref{fo}) in the dressed quark state. 
The non vanishing contribution comes from the quark-gluon state. We get,
\begin{eqnarray}
\Delta q_L(x,Q^2) && = \sum_{\sigma_1, \lambda_2}   ~\int d^2 \kappa^\perp
~(1-x)~
{\psi^{\uparrow}_{\sigma_1 \lambda_2}}^*(x, \kappa^\perp) 
(- i {\partial \over \partial \phi})
{\psi^{\uparrow}_{\sigma_1 \lambda_2}}(x, \kappa^\perp)  \nonumber \\
&& = -{\alpha_s \over 2 \pi} C_f ln{Q^2 \over \mu^2}  
(1-x) x^2 (1-x)^2
{ 1 \over 1-x} \Big [ { 1 \over x^2 (1-x)^2} - { 1 \over (1-x)^2} \Big ].
\end{eqnarray}
The first (second) term inside the square bracket arises from the 
state with gluon helicity +1 (-1).
Thus we have the quark orbital helicity contribution in the dressed quark
state
\begin{eqnarray}
\Delta q_L(x,Q^2) = - {\alpha_s \over 2 \pi} C_f ln{Q^2 \over \mu^2}   ~(1-x)
(1+x).
\label{qoh}
\end{eqnarray}
Similarly we get the gluon orbital helicity distribution defined in eq.
(\ref{go}) in the dressed
quark state  
\begin{eqnarray}
\Delta g_L(1-x,Q^2) && = \sum_{\sigma_1, \lambda_2}   ~\int d^2 \kappa^\perp
~~x~
{\psi^{\uparrow}_{\sigma_1 \lambda_2}}^*(x, \kappa^\perp) 
(- i {\partial \over \partial \phi})
{\psi^{\uparrow}_{\sigma_1 \lambda_2}}(x, \kappa^\perp)  \nonumber \\
&&= -{\alpha_s \over 2 \pi} C_f ln{Q^2 \over \mu^2}  ~ x
(1+x).
\label{goh}
\end{eqnarray}
We note that the helicity is conserved at the quark gluon vertex. 
For the initial quark of zero transverse momentum, total helicity of the
initial state is the intrinsic helicity of the initial quark, namely, $+ { 1
\over 2}$ in our case.
Since we have neglected quark mass effects, the final quark also has
intrinsic helicity $+ { 1\over 2}$. Thus total helicity conservation implies
that the contributions from gluon intrinsic helicity and quark and gluon
internal orbital helicities have to cancel. This is readily verified using eqs.
(\ref{gih}), (\ref{qoh}), and (\ref{goh}).

>From eqs. (\ref{gih}), (\ref{qoh}) and (\ref{goh}) we extract the relevant splitting
functions. The splitting functions are 
\begin{eqnarray}
P_{SS(gq)}(1-x) ~ &&= C_f~(1+x) ,\nonumber \\
P_{LS(qq)} (x) ~ &&= -~C_f~(1-x^2), \nonumber \\
P_{LS(gq)}(1-x) ~ &&= -C_f ~ x~(1+x).
\end{eqnarray}
We define the anomalous dimension $ A^n = \int_0^1 dx x^{n-1} P(x)$.
The anomalous dimensions are given by 
\begin{eqnarray} 
A^n_{SS(gq)} = C_f ~ { n+2 \over n(n+1)}, ~~
A^n_{LS(qq)} = -~ C_f ~ { 2 \over n(n+2)}, ~~ A^n_{LS(gq)} = -~ C_f ~{ n+4
\over n (n+1) (n+2)}.
\end{eqnarray}
These anomalous dimensions agree with those given in the recent work of 
H\"{a}gler and Sch\"{a}fer\cite{sch}.
\section { Verification of helicity  sum rule}
Helicity sum rule for the fermion target is given by
\begin{eqnarray}
{ 1 \over {\cal N}} \langle PS \mid \Big [ J^3_{q(i)} + J^3_{q(o)} +
J^3_{g(i)} + J^3_{g(o)} \Big ] \mid PS \rangle = \pm { 1 \over 2}.
\end{eqnarray}
For boson target RHS of the above equation should be replaced by the
corresponding helicity.

Here we verify the correctness of our definitions of distribution functions
in the context of
helicity sum rule for a dressed quark as well as
a dressed gluon target perturbatively. For simplicity, we take the external
transverse momenta of the target to be zero so that there is no net angular
momentum associated with the center of mass of the target. Using the field expansions,
given in Eqs.(\ref{psi}) and (\ref{ap}),
we have,
\begin{eqnarray}
J^3_{f(o)} && = i \sum_s \int {dk^+ d^2 k^\perp \over 2 (2 \pi)^3 k^+} \Bigg
[
b^\dagger(k,s) \Big [ k^2 { \partial \over \partial k^1} - k^1 { \partial
\over \partial k^2} \Big ] b(k,s)
+ d^\dagger(k,s) \Big [ k^2 { \partial \over \partial k^1} - k^1 { \partial
\over \partial k^2} \Big ] d(k,s) \Bigg ], \nonumber \\
J^3_{f(i)} && = {1 \over 2} \sum_\lambda \lambda
\int {dk^+ d^2 k^\perp \over 2 (2 \pi)^3 k^+} \Bigg
[b^\dagger(k,\lambda) b(k,\lambda)+
 d^\dagger(k,\lambda) d(k,\lambda) \Bigg ], \nonumber \\
J^3_{g(o)} && = i \sum_{\lambda}  \int { dk^+ d^2 k^\perp \over 2 ( 2 \pi)^3
k^+} a^\dagger (k, \lambda) \Big [ k^2 {\partial \over \partial k^1} - k^1
{\partial \over \partial k^2} \Big ] a(k, \lambda), \nonumber \\
j^3_{g(i)} && =  \sum_{\lambda}  \lambda \int { dk^+ d^2 k^\perp \over 2 ( 2 \pi)^3
k^+} a^\dagger (k, \lambda)  a(k, \lambda).
\end{eqnarray}
For a dressed quark target having helicity $+{1\over2}$ we get,
\begin{eqnarray}
{1\over {\cal N}}\langle P , \uparrow \mid J^3_{f(i)} \mid P, \uparrow
\rangle_q = && \int dx  \Big[ {1\over 2}\delta(1-x) + {\alpha\over 2\pi}C_f
ln{Q^2\over \mu^2}\big[ {1+x^2 \over (1-x)_+}
+{3\over2}\delta(1-x)\big]\Big]\nonumber\\
= &&~~~{1\over2}\nonumber\\
{1\over {\cal N}}\langle P , \uparrow \mid J^3_{f(o)} \mid P, \uparrow
\rangle_q = && - {\alpha\over 2\pi}C_f
ln{Q^2\over \mu^2}\int dx   ~(1-x)~(1+x)\nonumber\\
{1\over {\cal N}}\langle P , \uparrow \mid J^3_{g(i)} \mid P, \uparrow
\rangle_q = &&~~~  {\alpha\over 2\pi}C_f
ln{Q^2\over \mu^2}\int dx  ~ (1+x)\nonumber\\
{1\over {\cal N}}\langle P , \uparrow \mid J^3_{g(o)} \mid P, \uparrow
\rangle_q = && - {\alpha\over 2\pi}C_f
ln{Q^2\over \mu^2}\int dx ~ x ~(1+x).
\end{eqnarray}
Adding all the contributions, we get,
\begin{equation}
{1\over {\cal N}}\langle P , \uparrow \mid J^3_{f(i)}
+J^3_{f(o)}+J^3_{g(i)}+J^3_{g(o)}
\mid P, \uparrow
\rangle_q =~~~{1\over2}.
\end{equation}
For a dressed gluon having helicity $+1$, the corresponding expressions 
are worked out to be the
following.
\begin{eqnarray}
{1\over {\cal N}}\langle P , \uparrow \mid J^3_{f(i)} \mid P, \uparrow
\rangle_g = &&~~~0\nonumber\\
{1\over {\cal N}}\langle P , \uparrow \mid J^3_{f(o)} \mid P, \uparrow
\rangle_g = &&  {\alpha\over 2\pi}N_fT_f
ln{Q^2\over \mu^2}\int dx ~[x^2+ (1-x)^2]\nonumber\\
{1\over {\cal N}}\langle P , \uparrow \mid J^3_{g(i)} \mid P, \uparrow
\rangle_g = &&~~~ \psi_1^*\psi_1 \nonumber\\
=&&~~~1-  {\alpha\over 2\pi}N_fT_f
ln{Q^2\over \mu^2}\int dx~  [x^2+ (1-x)^2]\nonumber\\
{1\over {\cal N}}\langle P , \uparrow \mid J^3_{g(o)} \mid P, \uparrow
\rangle_g = &&~~~0 
\end{eqnarray}
Adding all the contributions, we get,
\begin{equation}
{1\over {\cal N}}\langle P , \uparrow \mid J^3_{f(i)}
+J^3_{f(o)}+J^3_{g(i)}+J^3_{g(o)}
\mid P, \uparrow
\rangle_g =~~~1.
\end{equation}
Note that in evaluating the above expression, we have used the
Fock-expansion of the target states. For the dressed quark we have used
Eq.(\ref{dsqs}), while for gluon we have used similar expansion but ignored
two-gluon Fock sector for simplicity.   

\section{summary, conclusions and discussion}
We have presented a detailed analysis of the
light-front helicity operator (generator of rotations in the transverse
plane) in QCD.  
We have explicitly shown that,
the operator constructed from manifestly gauge
invariant, symmetric energy momentum tensor in QCD, in the gauge $A^+=0$,
and after the elimination of constraint variables, is equal to the naive
canonical form of the light-front helicity operator plus surface terms. In
the topologically trivial sector, we can legitimately require the dynamical
fields to vanish at the boundary. This eliminates the residual gauge degrees
of freedom and removes the surface terms. 

Next, we have defined non-perturbative quark and gluon orbital helicity 
distribution functions as Fourier
transform of forward hadron matrix elements of appropriate bilocal operators
with bilocality only in the light-front longitudinal space.
We have calculated these distribution functions by replacing the hadron
target by a dressed parton providing all the necessary details. 
>From these simple calculations we have 
illustrated the utility of the newly defined distribution functions in the 
calculation of
splitting functions and hence anomalous dimensions in perturbation theory. 
We have also verified the helicity sum rule explicitly to the first non-trivial
order in perturbation theory. 
 
Lastly, in an appendix, we have compared and contrasted the expressions 
for internal
orbital helicity in non-relativistic and light-front (relativistic) cases. 
Our calculation shows that the role played by particle masses in the internal
orbital angular momentum in the non-relativistic case is replaced by the
longitudinal momentum fraction in the relativistic case.
Although four terms appear in the expression of $L_3$ for
individual particles in two body system, only the term proportional to the
total internal $L_3$ contributes due to transverse boost invariance of the
multi-parton wave-function in light-front dynamics. We also note the occurrence 
of the longitudinal
momentum fraction $x_2$($x_1$) multiplied by the total internal $L_3$ in the 
expressions
of $L_3$ for particle one(two). This explains why one needs to take first moment with
respect to $x$ as well as $(1-x)$ for the respective distributions in
obtaining the helicity sum rule\cite{ji}.

Our explicit demonstration that the  operator 
constructed from manifestly gauge
invariant, symmetric energy momentum tensor in QCD, in the gauge $A^+=0$,
and after the elimination of constraint variables and residual gauge freedom,
is equal to the naive
canonical form of the light-front helicity operator is facilitated by the
fact that in light-front theory only transverse gauge fields are dynamical
degrees of freedom. The conjugate momenta (color electric fields) are
constrained variables in the theory. Thus we 
were able to show explicitly that the
resulting gauge fixed operator is free of interactions.  
The question naturally arises as
to whether this result is valid in other gauges also. Several years ago, in
the context of magnetic monopole solutions, 
it has been shown\cite{cgw} that in
Yang-Mills-Higgs system, quantized in the axial gauge $A_3=0$ using the
Dirac procedure, the angular momentum operator constructed from manifestly
gauge invariant symmetric energy momentum tensor differs from the canonical
one only by surface terms. 
In the study of QCD in $A_3 =0$ gauge, it has been shown\cite{bg} that
in the presence of surface terms, Poincare algebra holds only in the
physical subspace.  
The situation in $A^0=0$ gauge or
in covariant gauges where unphysical degrees of freedom are present is to be
investigated. Another interesting problem to be studied 
is the helicity conservation in
the topologically non-trivial sector of QCD and its implications, if any,
for deep inelastic scattering.   

\acknowledgments

A.H would like to thank Wei-Min Zhang for useful communications.

\appendix
\section{Internal orbital helicity: Non-relativistic versus light-front
(relativistic) case}
First we address the non-relativistic versus the light-front case.
We need to decompose the total orbital angular momentum of a composite
system as a sum of the orbital angular momentum associated with internal motion and
the orbital angular momentum associated with the center of mass motion. We
are interested only in the former and not in the latter.  
For illustrative purposes, consider a two
body system consisting of two particles with masses $m_1$ and $m_2$ and
momenta ${\bf k_1}$ and ${\bf k_2}$. Let ${\bf P}$ denote the total momentum.
In the non-relativistic case, let ${\bf q}$ denote the relative
momentum, i.e., $ {\bf q} = { m_2 {\bf k_1}  - m_1 {\bf k}_2 \over m_1 +m_2}$. 
It is well-known\cite{chengli} that the contribution of particle
one (two)
to the third component of internal orbital angular momentum is given by
\begin{eqnarray}
L^3_{1(2)} = i {m_{2(1)} \over m_1 + m_2} \Big [ q^2 {\partial \over \partial
q^1} - q^1 {\partial \over \partial q^2} \Big]. \label{nrl3}
\end{eqnarray} 

Next consider the light-front case. 
Let $k_1=(k_1^+,k_1^\perp)$ and $k_2=(k_2^+,k_2^\perp)$ denote the
single particle momenta and  $P=(P^+,P^\perp)$ denote
the total momentum of the two particle system, i.e.,
$k_1^{+,i}+k_2^{+,i}=P^{+,i}$. Light-front kinematics allows us to introduce
boost-invariant internal transverse momentum $q^\perp$ and longitudinal
momentum fraction $ x_i$ by
\begin{eqnarray}
k_1^\perp = q^\perp + x_1 P^\perp, ~~k_1^+ = x_1 P^+, ~~~~
k_2^\perp = -q^\perp + x_2 P^\perp, ~~k_2^+ = x_2 P^+.
\end{eqnarray}
we have $x_1 + x_2 = 1$ and $ q^\perp = x_2 k_1^\perp - x_1 k_2^\perp$.
For the first particle, we have
\begin{eqnarray}
L_{1} && = i \Big [ k_1^2 { \partial \over \partial k_1^1} - k_1^1 { \partial
\over \partial k_1^2} \Big ] \nonumber \\
&& = ix_2 \Big [ q^2 { \partial \over \partial q^1} - q^1 { \partial \over
\partial q^2} \Big ] + 
i x_1 \Big [ P^2 { \partial \over \partial P^1} - P^1 { \partial \over
\partial P^2} \Big ]  \nonumber \\
&& ~~~~ +~i x_1 x_2 \Big [ P^2 { \partial \over \partial q^1} - P^1 { \partial \over
\partial q^2} \Big ] +
i \Big [ q^2 { \partial \over \partial P^1} - q^1 { \partial \over
\partial P^2} \Big ] .
\end{eqnarray} 
For the second particle, we have
\begin{eqnarray}
L_2 && = i \Big [ k_2^2 { \partial \over \partial k_2^1} - k_2^1 { \partial
\over \partial k_2^2} \Big ] \nonumber \\
&& = ix_1 \Big [ q^2 { \partial \over \partial q^1} - q^1 { \partial \over
\partial q^2} \Big ] + 
i x_2 \Big [ P^2 { \partial \over \partial P^1} - P^1 { \partial \over
\partial P^2} \Big ]  \nonumber \\
&& ~~~~-~i x_1 x_2 \Big [ P^2 { \partial \over \partial q^1} - P^1 { \partial \over
\partial q^2} \Big ] -
i \Big [ q^2 { \partial \over \partial P^1} - q^1 { \partial \over
\partial P^2} \Big ] .
\end{eqnarray} 
Total orbital helicity 
\begin{eqnarray}
L= L_1 + L_2 = i \Big [ q^2 { \partial \over \partial q^1} - q^1 {\partial \over
\partial q^2}\Big ] + i \Big [ P^2 { \partial \over \partial P^1 } - P^1 
{\partial \over \partial P^2} \Big ].
\end{eqnarray}
Thus we have decomposed the total orbital helicity of a two particle system
into internal orbital helicity and the orbital helicity associated with 
the "center of mass motion".

Note that the internal orbital helicity carried by particle one is the total
internal helicity multiplied by the longitudinal momentum fraction carried
by particle two and vice versa. This factor can be understood 
by comparison with the
situation in non-relativistic dynamics and recalling the close analogy
between Galilean relativity and light-front dynamics in the transverse
plane. 
In non-relativistic two-body problem, the center of mass coordinate is
defined by $ \stackrel{\rightarrow}{R} = {m_1 \stackrel{\rightarrow}{r_1} +
m_2 \stackrel{\rightarrow}{r_2} \over m_1 + m_2}$. The generator of Galilean
boost is $ \stackrel{\rightarrow}{B} = - \sum_i m_i
\stackrel{\rightarrow}{r_i}$.
 Thus in non-relativistic dynamics, $ \stackrel{\rightarrow}{R} =
- {\stackrel{\rightarrow}{B} \over M}$ with $ M= m_1 + m_2$.
In light-front dynamics, the variable analogous to $ B^\perp$ is $E^\perp$,
the generator of transverse boost and the variable analogous to $M$ is   
$P^+$. Thus in light-front theory, the transverse center of mass coordinate 
$ R^\perp = {\sum_i k_i^+ r_i^\perp \over \sum_i k_i^+} = x_1 r_1^\perp +
x_2 r_2^\perp$. Thus we recognise that instead of ${m_2 \over m_1+ m2}$ 
(${m_1 \over m_1+m_2})$ in non-relativistic theory, $x_2$ ($x_1$) appears in
light-front theory. 

By  comparing light-front (relativistic) and non-relativistic cases, 
we readily see that the
role played by particle masses in individual contributions to the third
component of internal
orbital angular momentum in non-relativistic dynamics is replaced by
longitudinal momentum fractions in relativistic (light-front) theory. 
This also shows that
the physical picture of the third component of internal orbital angular
momentum is drastically different in non-relativistic and relativistic cases.
We stress that it is only the latter, in which parton masses do not appear at
all, that  is of relevance to the nucleon
helicity problem. Lastly, we emphasize that it is 
the transverse boost invariance in light front dynamics that makes 
possible the separation of dynamics associated with the center of mass 
and the internal dynamics. 
{\it In equal-time relativistic theory, this separation
cannot be achieved at the kinematical level since boosts are dynamical}.
 

\end{document}